\documentclass[epjST]{svjour}
\usepackage{amsmath}
\usepackage{geometry}
\usepackage{amssymb}
\usepackage[usenames,dvipsnames]{color}
\usepackage{graphics}
\usepackage{graphicx}
\usepackage{subfigure}

\begin{document}
\title{Socio-Inspired ICT}
\subtitle{Towards a Socially Grounded Society-ICT Symbiosis}
\author{Alois Ferscha\inst{1}\fnmsep\thanks{\email{ferscha@pervasive.jku.at}}
\and Katayoun Farrahi\inst{2}
\and Jeroen van den Hoven\inst{3}
\and David Hales\inst{4}
\and Andrzej Nowak\inst{5}
\and Paul Lukowicz \inst{6}
\and Dirk Helbing\inst{7}}
\institute{University of Linz (JKU), Inst. f. Pervasive Computing, Altenberger Strasse 69, 4040 Linz, Austria
\and University of Linz (JKU), Inst. f. Pervasive Computing, Altenberger Strasse 69, 4040 Linz, Austria
\and Philosophy Section, Delft University of Technology, Jaffalaan 5, P.O. Box 5015, 2600 GA
Delft, The Netherlands
\and The Open University, London, UK
\and Department of Psychology, University of Warsaw, 00-183 Warsaw, Poland Stawki 5/7, Poland
\and DFKI, Trippstadter Straße 122, D-67663 Kaiserslautern, Germany
\and ETH Z¨urich, Clausiusstrasse 50, 8092 Z¨urich, Switzerland
}
\abstract{
Modern ICT (Information and Communication Technology) has developed a vision where
the ``computer'' is no longer associated with the concept of a single device or a
network of devices, but rather the entirety of situated services originating in
a digital world, which are perceived through the physical world. It is observed that
services with explicit user input and output are becoming to be replaced by a computing landscape
sensing the physical world via a huge variety of sensors, and controlling it via
a plethora of actuators. The nature and appearance of computing devices is changing
to be hidden in the fabric of everyday life, invisibly networked, and omnipresent, with
applications greatly being based on the notions of context  and knowledge.
Interaction with such globe spanning, modern ICT systems  will presumably be more
implicit, at the periphery of human attention, rather than explicit, i.e. at the focus of
human attention.
\\
\\
Socio-inspired ICT assumes that future, globe scale ICT systems should be
viewed as social systems. Such a view challenges research to identify and
formalize the principles of interaction and adaptation in social systems, so
as to be able to ground future ICT systems on those principles. This position paper therefore is concerned with the
intersection of social behaviour and modern ICT, creating or recreating social conventions
and social contexts through the use of pervasive, globe-spanning, omnipresent and participative ICT.
} 

\maketitle

\section{The Rise of Aware ICT}

Modern ICT, building on the ever progressing miniaturization of technology
(processing, storage, communication) as well as at the ever growing
globe spanning networks, has postulated to invisibly integrate technology
into everyday objects like tools, appliances, objects of everyday use, and
environments like offices,
homes and cars in such a way, that these objects turn into
``\textit{smart things}'' or ``\textit{smart environments}''.
Built with networked embedded systems technology, such ``smart'' things and environments
have become increasingly interconnected, diverse and heterogeneous, and
together with IP networking technology have created a whole new generation
of ICT as we see it today
(e.g the ``\textit{Internet-of-Things}'',  ``\textit{Smart Buildings, Cars, Cities}'',
``\textit{Smart Grids}'', even the ``\textit{Smart Planet}'').
Only networking and communication capabilities, however,
cannot make things and environments appear ``smart'',
unless coping with the challenge of an operative, and semantically meaningful interplay
among each other.

One approach to address the challenge of ``smart'' ICT has been to
design and implement systems able to manage themselves in a more or
less autonomous way, with little or no human interaction. While self-management
stands for the ability of single smart thing to describe itself, to select
and use adequate sensors to capture information describing its context,
self-organizing stands for the ability of a group of possibly heterogeneous
peers to establish a spontaneous network based on interest, purpose or
goal, and to negotiating and fulfilling a group goal. Self-management relates to
an individual smart thing, and concerns adaptation to changing individual
goals and conditions at runtime, while self-organization relates to whole
ensembles of smart things, and concerns adaptation in order to meet group
goals.

\begin{figure}[h]
  	\centering
	\includegraphics[width=0.9\textwidth]{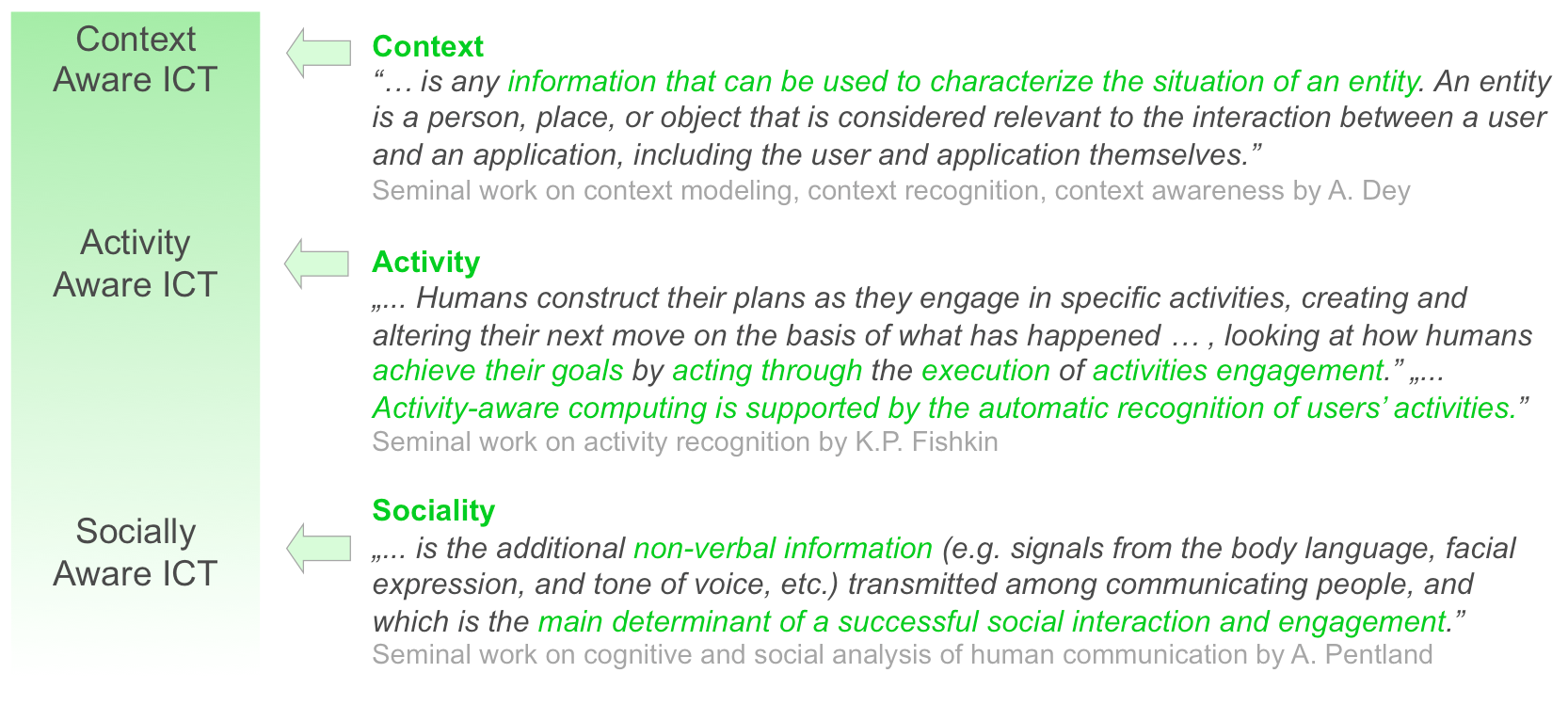}
	\caption{The Evolution of Aware ICT Systems}
   \label{fig:awareICT}
\end{figure}

A prerequisite for self-management, even more so for self-organization
is the ability to autonomously perceive, recognize, and even anticipate phenomena
and their consequences, i.e. being ``\textit{aware}''. 
Early signs ``\textit{aware ICT}'' have been observed by contributions from
Pervasive and Ubiquitous Computing
over the past two decades, starting with systems being aware about the physical situation
they are operated in (``\textit{context aware ICT}'')\cite{schilit_94}, and later on with systems being aware about the
user and his activities (``\textit{activity aware ICT}'')\cite{philipose2004inferring,roggen2011activity} (see Fig.  \ref{fig:awareICT}).
More recent trends tend to make ICT aware about social (``\textit{socially aware ICT}'')\cite{pentland_05socially,lukowicz2012context},
emotional (``\textit{emotion aware ICT}'')\cite{barry2005motion} and even cognitive aspects  (e.g.,
``\textit{attention aware ICT}'')\cite{ferschaindividual}. We look at this evolution in more detail.

\subsection{From Context Aware ICT to Socio-Technical Fabric}

Quoting from Weiser's (1991) vision ``\textit{The most profound technologies are those that
disappear. They weave themselves into the fabric of every day life, until they are
indistinguishable from it}'' \cite{Weiser1999} conveys the most common understanding of the origins
of a computer science research branch today known as Pervasive and Ubiquitous Computing (PUC).
Weiser's seminal vision was pathbreaking, and still represents the corner stone for
what might be referred to as a first generation of  research, aiming towards embedded, hidden, invisible, but
networked ICT systems. This first generation definitely gained from the technological
progress momentum (miniaturization of electronics, gate packaging), and was driven
by the upcoming availability of technology to connect literally everything to everything
(\textbf{Connectedness}, Late Nineties), like wireless communication standards and the
exponentially growing internet. Networks of systems emerged, forming
communication clouds of miniaturized, cheap, fast, powerful, wirelessly connected,
``always on'' systems, enabled by the massive availability of miniaturized computing,
storage, communication, and embedded systems technologies. Special purpose computing
and information appliances, ready to spontaneously communicate with one another,
sensor-actuator systems to invert the roles of interaction from human to machine
(implicit interaction), and organism like capabilities (self-configuration, self-healing,
self-optimizing, self-protecting) characterize this generation.

The second generation PUC research inherited from the then upcoming sensor
based recognition systems, as well as knowledge representation and processing technologies
(\textbf{Awareness}, around the turn of the century), where research issues like e.g. context
and situation awareness, self-awareness, future-awareness or resource-awareness reshaped
the understanding of pervasive computing. Autonomy and adaptation in this generation
was reframed to be based on knowledge, extracted from low level sensor data captured
in a particular situation or over long periods of time (The respective ``epoch''
of research on \textbf{``context aware''} systems was stimulated by Schillit, Adams and Want
\cite{schilit_94}, and fertilized by the PhD work of Anind Dey \cite{dey_01}, redefining the term ``context''
as: ``\textit{\dots any information that can be used to characterize the situation
of an entity. An entity is a person, place, or object that is considered relevant
to the interaction between a user and an application, including the user and application
themselves."). } One result out of this course of research are autonomic systems\cite{kephart_03},
and later autonomic elements, able to capture context, to build up, represent
and carry knowledge, to self-describe, -manage, and --organize with respect to the
environment, and to exhibit behaviour grounded on ``knowledge based'' monitoring,
analysing, planning and executing were proposed, shaping \textbf{ecologies} of ICT
systems, built from collective autonomic elements interacting in spontaneous spatial/temporal
contexts, based on proximity, priority, privileges, capabilities, interests, offerings,
environmental conditions, etc.

Finally, a third generation of PUC is approaching, building upon connectedness
and awareness, and attempting to exploit the (ontological) semantics of systems,
services and interactions (i.e. giving \textbf{meaning} to situations and actions).
Such systems are often referred to as highly complex, orchestrated, cooperative and
coordinated ``Ensembles of Digital Artefacts'' (FP7 FET). An essential aspect of
such an ensemble is its spontaneous configuration towards a complex system, i.e.
a ``...\textit{ dynamic network of many agents (which may represent cells, species,
individuals, nations) acting in parallel, constantly acting and reacting to what
the other agents are doing where the control tends to be highly dispersed and decentralized,
and if there is to be any coherent behavior in the system, it has to arise from competition
and cooperation among the agents, so that the overall behavior of the system is the
result of a huge number of decisions made every moment by many individual agents}" \cite{castellani2009sociology}.

Ensembles of digital artefacts as compounds of huge numbers of possibly heterogeneous
entities constitute a future generation of systems to
which we refer  as \textbf{Socio-Technical Fabric}\cite{ferscha2011pervasive}, weaving social and
technological phenomena into the `fabric of technology-rich societies'.
Indications of evidence for such large scale, complex, technology rich societal settings
are facts like 10${}^{12}$ -10${}^{13}$ ``things'' or ``goods'' being traded in (electronic)
markets today, 10${}^{9}$ personal computer nodes and 10${}^{9}$ mobile phones on
the internet, 10${}^{8}$ cars or 10${}^{8}$ digital cameras with sophisticated embedded
electronics - even for internet access on the go, etc. Today's megacities approach
sizes of 10${}^{7}$ citizens. Already today some 10${}^{8}$ users are registered
on Facebook, 10${}^{8}$ videos have been uploaded to YouTube, like 10${}^{7}$ music
titles haven been labeled on last.fm, etc. Next generation research directions are
thus going away from single user, or small user group as addressed in the
first two generations, heading more towards complex socio-technical systems, i.e.
large scale to very large scale deployments of PUC and the respective concerns
on a societal level \cite{ziapervasive}.

\subsection{Key Future ICT Research Challenges Identified by the Scientific Community}

To better understand the trends and impacts of future ICT systems, as well as the research
challenges posed by them, we have conducted a large scale solicitation
initiative to pave future generation ICT research roadmaps\footnote{The
FP7 FET proactive project PANORAMA (FET proactive / Goal 8.3: Pervasive Adaptation)
picked up on the challenge of identifying the new trails of Pervasive Computing research,
involving some 240 of the most distinguished researchers in the field in a solicitation
process that lasted for about three years. The result of this process is manifested in the
Pervasive Adaptation Research Agenda Book (www.perada.eu/research-agenda), an
evolving document where the scientific community can contribute to and download
the latest version in a ready to print layouted format}. The voices raised by active
researchers in the field can be clustered according to the following research challenges
for future generation PUC \cite{ferscha201220}.

 \textbf{Autonomous Adaptation} The first category of challenges articulated by the scientific community addresses
on systems related research concerning the ability of a system to adapt to situation
changes based on an autonomous recognition and assessment of the situation, and to
``facilitate the survival of the system". Parallels can be identified to the ongoing
self-* systems research, but issues are raised that reach far into foreign domains
like neurology, immunology (e.g. systems developing their own internal self-image
to guide interaction with the externa, E. Hart), or environmental research (e.g.
systems self-optimizing their configuration with respect to. to environmental constraints like
carbon footprint or global energy, D. Moriandi). The study of symbiotic multi-body
organisms and systems with homeostatic abilities (e.g. danger perception or self/non-self
discrimination, J. Timmis) is proposed, much like the relief of mobility and spatial
coverage constraints in wireless sensor networks (e.g. collaborative ``cloud sensing"
with the robotic flying sensor network, P. Zhang).

 \textbf{Adaptive Pervasive Ensembles}
 Heterogeneous multipart systems provisioning services as an orchestrated
service ensemble are challenging the community in several concerns. First, the integrative
aspect of service components on a hardware and middleware layer and how they adapt
to achieve service stability (K. Herrmann), second configuration aspect on the wireless
communication topology layer (e.g. nanoscale communications and nanonetworks, O.
Baris Akan), but also at the layer of orchestrated human-technology interaction at
societal scale (socially interactive computing, P. Lukowicz, D. Helbing and St. Bishop).
This latter research challenge statement (backed by whole FuturICT project) is even
addressing globe-spanning, complex, dynamic systems, where ``\textit{adaptation could
range from reshuffling of resources (e.g. information sources, bandwidth, distributed
computing resources) to enable a better monitoring and management of emerging crisis
situations, over the mediation of interactions in and between communities, up to
emergency 'slow down and ask human' mechanisms, preventing the system from accelerating
crises and systemic failures}'' (P. Lukowicz). Symptomatically, like many of the recently
evolving research themes that build on todays capability to collect and analyze data
at a scale that may reveal patterns of behavior of whole societies or even mankind
(e.g. internet traffic, mobile telephony, automotive mobility, energy consumption,
etc.), this category attempts for a sustainable, reliable, stable, trustworthy and
inclusive ICT with human society, rather than an individual user as the target.

\textbf{Emergence and Evolvability}
Understanding the principles of growing ICT systems according to phenomena
like emergence (i.e. the way how complex systems and structures arise from the combination
and multiplicity of very simple components or interactions), or evolvability (the
idea of evolvable systems originated from early research in cybernetics, where evolve-ability
is known as ``the ability of a population to produce variants fitter than any yet
existing'') is considered a research challenge to cope with the seamless integration
of future technologies with already existing ICT infrastructures. Systems must
have the ability to adapt to spontaneous, unforeseen and even frequently changing
technological infrastructures, while preserving the capability of interfacing to
established technologies. Changes in system design paradigms from constructive to
evolvable (``\textit{It is natural that we ask ourselves if it is possible to start
with a minimal architecture and let the system grow and develop by itself, as an
answer to the environment demands and system's goals}." E. Costa), or instructive
(``\textit{by 'instructing' each tiny component from a network of components to increase
a specific benefit}", G. Persiona) and long-term ``\textit{self-developmental}"(S.
Kernbach) is where scientists identify need for research. The challenge to harmonize
-at least a temporal coexistence of- radical innovations with existing technology
is made clear by the example of 4D images (O. Bimber): color encoding only spatial
information of a scene in pixels of raster-displayed images (2D) could be enhanced
to also encode angular information, e.g. individual color for each emitting direction
of a pixel (4D). How would 4D light field photography/cameras and light field displays
coexist with traditional 2D imaging as a ICT enabling technology?

 \textbf{Societies of Artefacts}
 Conceivably, future ICT will be manifested by technology rich artefacts
(like tools, appliances, objects of every day use), and environments (like work and
home places, or sports and entertainment locations), cooperatively attempting service
provision with society-like behavior. Going beyond their capability to localize and
recognize other artefacts as well as humans and their intentions, societal artefacts
will form up to ``goal tribes'', i.e. ensembles of possibly complementing competencies,
to act in a sensitive, proactive, and responsive way according to the perceived and
anticipated needs, habits, and emotions of the users. While the social ability of
such artefacts is just the demanding prerequisite, the ability to form goal driven
interest communities according to societal models is the potential approach to harness
an ever increasing complexity of ICT. Coordinated goal oriented artifact communities
(``\textit{engineered to form societies, interact and compete with other ecologies,
collaborate with humans and develop their own methods of conception and social norms}",
A. Kameas) are supposed to be the ``interface'', via which humans will ultimately
be served. Research on the conception of artifact societies may supposedly inherit
from social and cognitive sciences, as it appears to be challenged to understand
the ``\textit{innovative intersection of norm-governed systems, voting algorithmics,
game theory, opinion formation, belief revision, judgement aggregation, and social
computational choice ..., as well as a formal characterization of socio-cognitive
principles of trust, forgiveness, and affect" }(J. Pitt).

 \textbf{Dependable Pervasive Systems}
 Particularly for ICT systems of very large deployment, non-functional
and quality-of-service system properties become prevalent over the pure services.
Beyond traditional dependable systems research (availability, reliability, fault
tolerance, maintainability), normative, ``self-regulating" system design approaches
are requested (a normative system refers to any distributed interaction system whose
behavior can be regulated by norms, e.g. norms to meet stability objectives or to
sustain certain utility levels) to be investigated. Much like individual human behavior
locally ``controlled" by (social, ethical, etc.) norms yields rational societies,
ICT artefacts could be organized as ensembles of "\textit{mission components
capable of assimilating and acting upon local intelligence 'on the fly'}" (J. Pitt).
Aside technological QoS criteria, more outreaching notions of dependability research
issues are proposed, like e.g. ``\textit{sustainable}" (``\textit{in terms of cost,
life-cycle management or energy efficiency}", St. Haller), ``\textit{socially meaningful}"
(P. Lukowicz), privacy preserving, or avoiding electronic waste and ``\textit{recyclability}"
(V. Namboodiri).

 \textbf{Pervasive Trust}
 Among the most prevalent, ubiquitously recognized, and meanwhile also socially
pressing research agenda items relate to the concerns humans might have in using
and trusting ICT. Well beyond the privacy and trust related research we see in
the Pervasive/Ubiquitous Computing research community already today, go the claims
for trust research in the context of information spread via the internet, in emergency
scenarios (e.g. advice-taking from technology or strangers), disaster management
(tsunamis, floodings, nuclear disasters, riots and civil commotion, etc., E. Mitleton-Kelly)
or risk analysis and crises forecasting (D. Helbing). The proposed working definition
of trust; ``... \textit{willingness to rely on another party and to take action in
circumstances where such action makes one vulnerable to the other party}" (E. Mitleton-Kelly)
already indicates its relation to risk (due to uncertainty), and vulnerability (due
to  the readiness to act). Inspired by recent results on the assessment of trust
users have in information delivered to ICT in particular situations (e.g. vibro-tactile
directional guidance in evacuation situations, www.socionical.eu) are upcoming questions
on how trust builds up (as a cognitive process from intuition, belief, experience/knowledge
and expectation), how to integrate trust building processes into ICT, how to
cope with distrust or lost trust, etc. The community agrees, that serious research
on Pervasive Trust cannot  survive without a formalisation of the trust related cognitive
capabilities and terms (experience, belief, expectation), and a foundational underpinning
with empirical evidence on trust mechanisms.

 \textbf{Human-Centric Adaptation}
 The standard phrase to approach a small-talk conversation among ICT
scientist is the question: ``Is it that human needs, capabilities or constraints shape
the design and emergence of PUC technology, or do humans adapt to technology designs
once it is deployed?''
Clearly, technology innovation and the processes of technology adoption by humans
is self-referential. Among the many examples throughout the evolution of ICT
which have made it to industrial mass products and/or commercial success, are smartphones
(the need to voice communicate vs. the joy of playful media and service access),
car navigation systems (the need for wayfinding vs. online traffic management) or
the internet of things (the need for identity management vs. total surveillance).
Opposed to a (yet vibrant) HCI research agenda, which usually assumes
users and systems as objects, and their interaction as subject of investigation,
researchers now try to explain adaptation based on how users ``understand" technology,
raising ``mental models" about how PUC systems work to become subject of investigation.
``Intelligibility" (coined as ``\textit{helping users to form an accurate mental model
about how to use an application}", A. Dey) and the design of ICT systems that lifelong
and ``\textit{continuously evaluate the degree of satisfaction (or frustration) they
elicit in users}" based on explicit but also implicit feedback (A. Roggen), that
allow users ``\textit{... to ask why did the system take a particular action, why
didn't it take a different specific action}" (A. Dey), that analyse physiological
dynamics of users (based on sensory input ``\textit{such as heartbeat, brain waves,
blood pressure, oxygen saturation, muscular activity, respiration, body temperature,
etc.}", J. Kantelhart) as a statistical time series, that ``\textit{truly understand
our mental and emotional situation and try to accommodate us}" (J. Healy), or that
trace and extract ``\textit{life patterns}" as indicators of lifestyles (M. Mamei)
- all in order to better understand how users ``understand" technology. Recent research
has coined the term ``\textit{scrutability}'' to underline the necessity for users
to be always able to inspect the knowledge of the system about them, and also to
help users to understand system failures and their causes. We could refer to intelligible
system designs also as ``respectful", i.e. respecting ``\textit{peoples' ability to
judge for themselves and be assisted by machines where needed. Respect for peoples'
desire for freedom of choice and be supported by automation and decision support
where appropriate" }(S. Spiekermann).

 \textbf{Socio-Technical Systems}
 A significant trend in next generation ICT research, already observed
taking its first steps, are investigations along the boundaries where technology
and social behavior interact. From the observation that success PUC technologies
(smartphones, mobile internet, autonomous driver assistance systems, social networks,
etc.) have radically transformed individual communication and social interaction,
the scientific community claims for ``\textit{new foundations for ... large-scale
Human-ICT organisms and their adaptive behaviours, also including lessons form applied
psychology, sociology, and social anthropology, other than from systemic biology,
ecology and complexity science.}" (F. Zambonelli). Moving research attention from
PUC for individual users towards the interplay of a complex, globe spanning,
dynamically changing ICT and societies of billions of users worldwide (``\textit{Socially
Interactive Computing}", P. Lukowicz) reveals the inadequacy of \textit{ceteris paribus} analysis:
at levels of such scale and complexity, behavioral phenomena cannot be explained
by investigations of influence quantities in isolation. Integrative views and research
methods have to be adopted to explain technology influenced social dynamics. The
community is challenged to extend the notion of ``\textit{context aware}" towards
``\textit{socially aware}" ICT \cite{pentland_05socially}. Remember, according to Dey, a system is called
context aware if it makes use of potentially any information that describes its situation.
In analogy, we would call a system socially aware, if it makes use of potentially
any information that describes its social habitus, i.e. any information which can
be inferred from all of its past and present social relations, social interactions
and social states. As for today, only a small subset of information constituting
social context can be captured via (technical) sensors (use patterns of social networking
tools, communication and mobility patterns captured by mobile devices, social apps,
calendar sharing, embedded geo tags, road pricing, financial transactions, etc.),
or is not conveniently accessible (hosted by public registries, state authorities,
service providers), but is available in principle, and has stimulated research domains
like computational social science \cite{lazer_09}.

 \textbf{Quality of Life} To support human beings towards a better life, the wellbeing of individuals,
as well as the welfare of societies are the quintessential prospect of PUC. Taking
all the technological capabilities and potentials, as well as the human desire for
wellbeing (``\textit{... a good condition of existence characterized by health, happiness,
and prosperity of individuals in relation to their inner and outer personal spheres}",
O. Mayora) together, several questions arise on how to reach a satisfactorily wellbeing
state via ICT. While lead and motivated by such humanistic principles towards
making the ``world a better place", it appears that the scientific community, at least
for the time being, has not yet found substantiated research questions. More are
there expressions of desirable world states (``\textit{Green World}", M. Ulieru) and
appreciable life styles (``\textit{freedom}", A. Schmidt, ``\textit{stress-free society}",
M. Mamei), reflecting what i would call ``realistic fiction", but awaiting more specificity,
structure and method.
\\

\noindent Some selected, yet indicative voices raised by the scientific community
towards the next generation ICT research challenges deserve to be highlighted:
\\

\small

\noindent \textbf{Intelligibility}\textit{ ``One particular
usability aspect of interest is }\textbf{\textit{intelligibility}}\textit{, helping
users to form an accurate mental model about how to use an application. This is important
for allowing users to understand how the application works and to be able to predict
what it will do in a future situation, and all of this will} \textit{impact adoption
and use.'' }(Anind K. Dey)

\noindent \textbf{Social and Cultural Knowledge}\textit{``Advanced data analysis tools will allow spotting
trends, observing their movement, their causes, and triggers. This platform, will
allow researchers to \dots  }\textbf{\textit{explore}}\textit{ }\textbf{\textit{social}}\textit{ and }\textbf{\textit{cultural}}\textit{ }\textbf{\textit{knowledge}}\textit{.
What do people }\textbf{\textit{believe}}\textit{? And how} \textit{people }\textbf{\textit{act}}."
(Adrian D. Cheok)
 \\

\noindent \textbf{Life-long
Satisfaction}\textit{ ``We believe that the next frontier in pervasive smart assistance
will be to devise systems capable of continuous - }\textbf{\textit{lifelong}}\textit{ - }\textbf{\textit{co-adaptation }}\textit{to
the user needs. \dots However, towards what should be the system adapted to, and
by what should adaptation be driven? We believe there is no better way than to be
guided by the }\textbf{\textit{satisfaction}}\textit{ of the user when he is interacting
with the system.'' }(Daniel Roggen)
\\

\noindent \textbf{Harvest
Ingenuity}\textit{ ``.. a new research approach in pervasive computing centering
on the investigation and development of human machines systems that increase the
freedom, utilize the power of communities, }\textbf{\textit{harvest}}\textit{ the }\textbf{\textit{ingenuity}}\textit{ of
a }\textbf{\textit{large}}\textit{ }\textbf{\textit{number}}\textit{ of }\textbf{\textit{independent}}\textit{ }\textbf{\textit{developers}}\textit{,
and develops technologies that address people's basic needs\dots '' }(Albrecht Schmidt)
\\

\noindent \textbf{Uncertainty}\textit{ ``Context-awareness
is woefully limited in our computing devices and they rarely do the "right thing"
or what we would prefer. We need to be able to teach them }\textbf{\textit{how}}\textit{ }\textbf{\textit{we}}\textit{ }\textbf{\textit{want}}\textit{ }\textbf{\textit{them}}\textit{ to
work for us\dots } \textit{What we need are the ability to tell if a user is interruptable,
what information they likely to need next, what work/play they might be engaged in,
and who might be engaged in it with them \dots  }\textbf{\textit{our models will
never cover all possible activities in which humans may engage}}\textit{.'' }(Gaetano
Borriello)
\\

\noindent \textbf{Thinking}\textit{ ``Building
on recent fMRI discoveries of common spatial patterns among subjects when thinking
of the same word, there are numerous projects processing brain signals in an} \textit{attempt
to understand }\textbf{\textit{what people }}\textit{are }\textbf{\textit{thinking}}\textit{.'' }(Daniel P. Siewiorek)
\\

\noindent \textbf{Cognitive
Adaptation}\textit{ ``One of the next grand challenges for adaptive pervasive computing
will be to make devices that truly }\textbf{\textit{understand}}\textit{ our }\textbf{\textit{mental}}\textit{ and }\textbf{\textit{emotional}}\textit{ }\textbf{\textit{situation}}\textit{ and
try to accommodate us." }(Jennifer Healey)
\\

\noindent \textbf{Knowledge
Accelerators}\textit{ ``We need to create a techno-socio-economic knowledge accelerator
-- a large scale multidisciplinary project that uses current and future ICT developments
to address the }\textbf{\textit{challenges of humanity }}\textit{involving natural
scientists and engineers.'' }(Dirk Helbing, Steven Bishop and Paul Lukowicz)
\\

\noindent \textbf{Knowledge
Self-Organisation}\textit{ ``\dots  devising the most effective }\textbf{\textit{mechanisms}}\textit{ }\textbf{\textit{for}}\textit{ }\textbf{\textit{knowledge}}\textit{ }\textbf{\textit{self-organisation}}\textit{ -
including knowledge creation, propagation and dissipation - to be used both at the
individual (in the workspace) and the global (in the information infrastructure)
levels \dots `` }(Andrea Omicini)
\\

\noindent \textbf{Life
Patterns}\textit{ ``\dots  technologies allow us - for the first time in history
- to collect large scale quantitative information about another fundamental realm
of nature: the }\textbf{\textit{daily}}\textit{ }\textbf{\textit{life}}\textit{ and }\textbf{\textit{daily}}\textit{ }\textbf{\textit{behavior}}\textit{ of
people.} ... \textit{One of the most interesting applications of this research considers
the "quality of life" and the "life style'' \dots  to contribute the creation of
"}\textbf{\textit{stress-free societies}}\textit{''. }(Marco Mamei)
\\

\noindent \textbf{Collective
Intelligence}\textit{The importance of emerging collective intelligence cannot be
denied, as it is the fact that pervasive computing technologies will make collective
intelligence so deeply embedded in our activities to make it impossible \dots  to
distinguish about what aspects of our ``}\textbf{\textit{intelligence}}\textit{" are
to be attributed to us as }\textbf{\textit{individuals}}\textit{, to us as member
of the world }\textbf{\textit{society}}\textit{, or to us as a organs of a continuous
and worldwide }\textbf{\textit{ICT-Social substrate}}\textit{. }(Franco Zambonelli)
\\

\noindent \textbf{Digital Formations}\textit{ ``
\dots  engineering of autonomous intelligent systems that co-exist with people in
real and synthetic environments - also referred to as "}\textbf{\textit{digital}}\textit{ }\textbf{\textit{formations}}\textit{"
or "}\textbf{\textit{digital}}\textit{ }\textbf{\textit{spaces}}\textit{'' \dots } \textit{engineered
to form societies, interact and compete with other ecologies, collaborate with humans
and develop their own methods of conception and social norms\dots "  }(Achilles Kameas)
\\

\noindent \textbf{Software Ecosystems}\textit{ ``A
world of highly interlinked pervasive devices, smart objects and smart environments
will only emerge if we succeed in }\textbf{\textit{unleashing}}\textit{ }\textbf{\textit{economic}}\textit{ and }\textbf{\textit{commercial}}\textit{ }\textbf{\textit{forces}}\textit{ that
will create a self-sustaining Pervasive Software Ecosystem that provides a playing
field for commercial (and non-commercial) software developers, providers, distributors,
vendors and end-users.'' }(Gerd Kortuem)
\\

\noindent \textbf{Behaviour
Specification}\textit{ ``We envision a system design methodology that relieves the
developer from }[''coding adaptation'']\textit{ \dots  one should be able to }\textbf{\textit{specify}}\textit{ the }\textbf{\textit{desired}}\textit{ global
system }\textbf{\textit{behavior}}\textit{ using appropriate high-level specification
languages. The pervasive system should then be endowed with an infrastructure to
develop adaptation strategies for its components such that the desired global behavior
is delivered across all possible situations.'' }(Friedemann Mattern)
\\

\noindent \textbf{Programming
Ensembles }\textit{``\dots  a big challenge is how to }\textbf{\textit{program such
populations in the large}}\textit{. For example, one would like to state high level
"suggestions" like "reduce your energy spending", "merge two populations", " please
elect a leader", "spread the information by an epidemic process", "increase the security
level" etc. and, ideally, the underlying population should be able to implement these
in a scalable (independent of current population size) and flexible way (e.g. choose} \textit{among a variety of routing methods).'' }(Paul Spirakis)
\\

\noindent \textbf{Digital
Ecosystems }\textit{``In what we refer to as '}\textbf{\textit{digital ecology}}\textit{'
theory and practice, research aims to understand and advance the interweaving of
humans and ICTs to create a world with social, physical, and cyber dimensions enabling
a kind of social network in which humans are }\textbf{\textit{not just 'consumers' }}\textit{of
data and computing applications \dots  } \textit{they are producers, 'players,' and
'inputs' whose interactions use the 'invisible hand' of the market as they interact
in complex, interdependent global-scale systems in areas such as energy production
and use, and neighborhood, district, city, and regional transport.'' }(Mihaela Ulieru)
\\

\noindent \textbf{Dependability }\textit{``
\dots  these teams have to be }\textbf{\textit{self-regulating}}\textit{, in terms
of a dynamic re-allocation of roles, tasks, priorities etc., which can be specified
as part of the normative system itself. A major challenge is to define }\textbf{\textit{dependability}}\textit{,
in terms of being able to meet specific organizational objectives and levels of} \textit{utility,
at the same time being able to withstand component-loss, network outage or overload,
and/or hostile behaviour.'' }(Jeremy Pitt)
\\

\noindent \textbf{Social
Values }\textit{``}\textbf{\textit{Respect}}\textit{ for peoples' ability to judge
for themselves and be assisted by machines where needed. Respect for peoples' desire
for freedom of choice and be supported by automation and decision support where appropriate.
And respect for fundamental human rights, such as privacy, security and safety.} \textit{A
key research area is thus how to build respect for }\textbf{\textit{humans' social
values}}\textit{ into the fabric of machines, to deepen our understanding of 'value
sensitive design'.'' }(Sarah Spiekermann)
\\

\noindent \textbf{Trustworthiness and Privacy }\textit{``There
is an obvious challenge in this personalization regarding the privacy of the collected
information: }\textbf{\textit{who is to store all this data, for how long, where,
and what is it used for?}}\textit{'' }(Mark Langheinrich)
\\

\noindent \textbf{Dual
Spatial Reality }\textit{``Problem states can more easily be transferred from the
real into the digital domain (by sensors) and the results of reasoning processes
of the digital domain can directly be transferred back into the real world (by actuators).
This tight connection between the digital and real world is what will lead to a }\textbf{\textit{Dual
Spatial Reality }}\textit{\dots `` }(Antonio Krueger)
\\

\noindent \textbf{Mobile
Augmented Reality }\textit{``The challenge is better connect remote people than with
a mobile phone employing }\textbf{\textit{context aware augmented reality}}\textit{.
Web 2.0 technologies have added to people's ability to stay connect with colleagues,
friends and family \dots  PC's \dots  do not scale down to smart phone form factors.
New technologies need to be investigated to overcome these issues, but in addition
take advance of the} \textbf{\textit{mobile nature of people}}\textit{\dots ``  }(Bruce
Thomas)
\\

\noindent \textbf{Smart
Material }\textit{``For pervasive systems, computing is material in two ways. First,
pervasive systems must intrinsically involve computing. Second, and more subtly,
the computing aspects of the system must be treated the same as any other }\textbf{\textit{material}}\textit{ that
affects the feel and behavior of an object.\dots  when computing is material, products
will have "}\textbf{\textit{smart patinas}}\textit{'', with their wear patterns determined
both physically and computationally.'' }(Tom Martin)
\\

\noindent \textbf{Meaningful
Applications }\textit{``We have struggled to enable }\textbf{\textit{large-scale
explorations }}\textit{of }\textbf{\textit{socially meaningful applications}}\textit{.
These applications include home health, elder care, and energy and resource monitoring
\dots `` }(Shwetak Patel)
\\

\noindent \textbf{Energy
Awareness}\textit{ ``\dots  an important issue that will have great impact on how
pervasive clouds will become is that of energy consumption. The development of }\textbf{\textit{energy-conscious}}\textit{ and }\textbf{\textit{power-aware}}\textit{ resource
allocation protocols for cloud computing systems will open up more opportunities
for the deployment of more pervasive technologies\dots `` }(Albert Zomaya)
\\

\noindent \textbf{Electronic
Waste}\textit{ ``Pervasive computing at scale via portable devices has social implications
in terms of }\textbf{\textit{electronic waste}}\textit{. For example, there are 4.2
billion mobile phones in use globally, with less than 3\% typically recycled according
to a study \dots Current mobile phones are replaced every 18-24 months, mainly to
obtain devices with better performance\dots `` }(Vinod Namboodiri)
\\

\normalsize

\section{Towards ICT Grounded on Social Principles}

Human beings are born dependent, and in constant need of support by others. When
growing up, humans do not gradually become independent of others, but rather become
interdependent. In the course of our lives we form many give-and-take relationships, building a healthy
interdependence with family, friends, communities, society and culture. \textit{``We
are, at our cores, social creatures. Affiliation is the strength that allows us to
join with others to create something stronger, more adaptive, and more creative than
any individual: the group"} (see Belonging to the Group, B.D. Perry,
2002). As Schopenhauer implies, the desire for \textbf{positive social relationships
is one of the most fundamental and universal of human needs}. This need has deep
roots in evolutionary history and exerts a powerful impact on contemporary human
psychological processes. 
Failure to satisfy
this need can have devastating consequences for psychological well-being. People
who lack positive relationships often experience loneliness, guilt, jealousy, depression,
and anxiety, higher incidence of psychopathology, 
and reduced immune system functioning. 
The psychological and sociological research in this areas has
observed that, given the strong need for social connection, the lack or weak perception
of \textbf{``social affiliation"} will fertilize two possible reactions: (i) antisocial---rather
than affiliative---responses to exclusion \cite{twenge2002social},
and (ii) an increase in motivation to build social bonds,
perhaps especially with new (and possibly more promising) social partners \cite{maner2007}.

The process of the social affiliation of a person in a societal community could be
affected by distortions and obstacles in case of individuals that live the social
marginalization or are at risk, because of one or more conditions of disadvantage:
personal disadvantages (e.g. physical and mental disability, psychological problems,
drug/alcohol addiction, being a prisoner or an ex-offender) and social disadvantages
(family in economic troubles, homeless, long-term unemployment). In these cases,
individuals need an external and professional support to foster the personal social
inclusion, through an individualized work directed to overcome the disadvantages
factors and enhance the personal and social abilities.

In our society, social services have taken up care to support social inclusion (or
re-integration), for example via personal caretakers. The personalized social assistance
provided is based on instruments and data coming (i) from him/her and his/her enlarged
family context (primary source), (ii) friends and peer-groups, or the neighborhood
(secondary source), and (iii) data records from societal authorities (schools records,
banking and insurance records, clinical records, police records, etc.). While the
primary and secondary sources are very important (they help the social intermediaries
to know the past and present life history of an individual and establish personal
contacts), they have also serious limitations: they originate from personal perceptions
and records, based on the individuals` biological senses (visual auditory, tactile)
and the ability to remember, forget and elaborate the social experiences - and thus
are prone to misinterpretation. Moreover, it is important to take into account that
the daily life is made of multiple factors affecting the ``social affiliation'' of
individuals as single entities and as part of larger societal bodies and the ``social
inclusion'' in the community. The biological senses (visual auditory, tactile)
and the ability to elaborate the social experiences are certainly important to feel
part of the physical and social environment around, but they need to be read together
with another phenomenon of humans that is very important: the ``social awarenes",
based on the ``social sense", i.e. an additional human sense that helps people to
perceive the ``social" aspects of the environment, allowing to sense, explore and
understand the social context.

The awareness of a social sense, and the ability to focus on, understand and express
without personal distortions what the social context means for the individual and
how it can influence a behaviour, could be the key-factors for the decisive improvement
of the individualized social inclusion interventions of disadvantaged people by social
intermediaries (i.e. public and private social services providers and their practitioners,
associations representative of disadvantaged groups). In this respect, new
technologies and ICTs could contribute to gather direct, accurate and intelligible
information on how the user experiences social relations: this innovation could
have very powerful effects for the enhancement of the vulnerable people social inclusion,
because it would complete the set of data and information about the user personal
profile provided by the traditional sources, improving the effectiveness of the inclusion
personalized intervention provided by social practitioners.

In summary, the rational underlying socio-inspired ICT research is that new technologies
 could contribute to gather direct, accurate and intelligible information on
how the user experiences social relations. This in turn could have very powerful
effects on $(i)$ the design of future ICT systems per se,
$(ii)$ the interaction principles between humans and ICT,
but also $(iii)$ the enhacement
of a flourishing symbiosis of society and ICT overall. Towards the identification
of research issues related to the potentials of underpinning new generation
ICT on social principles, a variety of terms have been introduced and studied
by the scientific community. Among them are:
\\

\textbf{Social Context} in computing is often used as a term commonly referring to
the people, groups, and organizations an individual is interacting with \cite{schuster_12}. There are some
variations of this definition which are more specific and can additionally differentiate between social versus non-social context. Pentland \cite{pentland_05socially} argues about the importance of social context and describes it as the additional non-verbal information (e.g. signals from the body language, facial expression, and tone of voice, etc.) transmitted among communicating people, and which is the main determinant of a successful social interaction and engagement. An application of social context in this respect, for example, can be the finding of new contacts and the integration of remote users in conversations. There has been a lot of progress in the domain of social signal processing, which builds on this definition by Pentland and focuses on non-verbal behavior in automatically recognizing social signals and social context \cite{vinciarelli_12,vinciarelli_09}. Groh et al.  \cite{groh_applicationsfor} define social context as \textit{``all social relations, social interactions and social situations which are directly related to or confined to small time-intervals and space-regions around the present time or present location of a person''}. Schuster et al. \cite{schuster_12} consider social context from a pervasive perspective, where they define
\textit{``pervasive social context of an individual as the set of information that arises out of direct or indirect interaction with people carrying sensor-equipped pervasive devices connected to the same social network service. … It comprises the explicit links, profile information and activities of people within the social graph, the joint sensor information of the pervasive devices as well as implicit information that can be inferred by combining the two.''} Building on Dey's definition of context (see above) \cite{dey_01} ), social context is any information of a social nature, including both non-verbal and verbal, transmitted among communicating people, that can be used to characterize the situation of an entity.

\textbf{Social Sense} The biological senses (visual auditory, tactile) and the ability to elaborate the social experiences are certainly important to feel part of the physical and social environment around, but they need to be read together with another phenomenon of humans that is very important: the "social awareness", based on the "social sense", i.e. an additional human sense that helps people to perceive the "social" aspects of the environment, allowing to sense, explore and understand the social context.

\textbf{Social Awareness} In \cite{pentland_05socially} Pentland identifies one key weakness in today's technology, the fact that they are socially ignorant. He writes
\textit{"Technology must account for [the fact that people are social animals], by recognizing that communication is always socially situated and that discussions are not just words but part of a larger social dialogue."} Additionally, initial steps have been taken by Pentland's research group to develop three socially aware platforms that objectively measure several aspects of social context, including nonlinguistic social signals. This recognition of both the larger social dialogue and the social context of a communication is what we define as social awareness. In the literature, there is no formally accepted definition of social awareness. Nowak and Conte \cite{} identify social awareness as \textit{``... the capacity to model ongoing social processes, structures and behavioural patterns...''}. We build on this and define social awareness as a property that may enable technology to automatically and objectively recognize ongoing social processes, social context, social structures and behavioural patterns. Social awareness extends context awareness by considering the social dimension as the dominant feature of interest.

\subsection{Mining for Social Context}

Modern ICT, as for example smartphones, have started to continuously sense movements, interactions, and potentially other clues about individuals, thus also about society as a whole. Data continuously captured by hundreds of millions of personal devices around the world, promises to reveal important behavioral clues about humans in a manner never before possible. Eagle and Pentland~\cite{eagle_09} performed the first Reality Mining data collection~\cite{eagle_05thesis}, which was named by MIT Technology Review as ``one of the 10 technologies most likely to change the way we live''~\cite{mittechnologyreview_website}.

Research using mobile phone data has mostly focused on location-driven analysis, more specifically, using Global Positioning System (GPS) data to predict transportation modes \cite{patterson_03,reddy_08}, to predict user destinations~\cite{krumm_06}, or paths~\cite{akoush_07}, to discover a user's stay regions (or places of interest)~\cite{montoliu_10}, and to predict daily step count~\cite{sohn_06}. Other location-driven tasks have made use of Global System for Mobile Communications (GSM) data for indoor localization~\cite{otsason_05} or WiFi for large-scale localization~\cite{letchner_05}. There are several works related to activity modeling from location-driven phone sensor data. CitySense~\cite{loecher_09} is a mobile application which uses GPS and WiFi data to summarize ``hotspots" of activity in a city, which can then be used to make recommendations to people regarding, for example, preferred restaurants and nightclubs~\cite{sensenetworks}. Liao et al.~\cite{liao_06} use GPS data traces to label and extract a person's activities and significant places. Their method is based on Relational Markov Networks. The BeaconPrint algorithm~\cite{hightower_05} uses both WiFi and GSM to learn the places a user goes and detect if the user returns to these places.

There has also been some previous work pertaining to modeling users and their mobile phone usage patterns. Eagle and Pentland~\cite{eagle_09eigenbehaviors} use Principle Component Analysis (PCA) to identify the main components structuring daily human behavior. The main components of human activities, which are the top eigenvectors of the PCA decomposition are termed {\it eigenbehaviors}. To define the daily life of an individual in terms of eigenbehaviors, the top eigenbehaviors will show the main routines in the life of a group of users, and the remaining eigenbehaviors describe the more precise, non-typical behaviors in individuals' or the group's lives. Farrahi and Gatica-Perez~\cite{farrahi_11acmtist,farrahi_10themes} have proposed a method to discover routines of users by modeling socio-geographic cues using topic models. This methodology of discovering Reality Mining-based behaviors was then extended to determine the similarities and differences between groups of people in a computational social science experiment. The experiments are performed on a political opinion dataset, where the authors use their approach to determine the similarities and differences in the daily routines of individuals who changed political opinions versus those that do not~\cite{madan_11}. Further, Do and Gatica-Perez~\cite{do_10} recently presented an analysis of application usage in smartphones, for the purpose of user retrieval. Similarly, Verkasalo et al.~\cite{verkasalo_10} studied the reasons and motivation behind using applications across users and non-users.

Building on the many previous works based on the sociometer~\cite{choudhury_03sociometer}, which is a wearable sensing device capable of sensing the amount of face-to-face interaction, conversational time, physical proximity to other people, and physical activity levels, mobile phones have been programmed to capture non-linguistic speech attributes~\cite{lu_09,madan_06}. These non verbal speech features have been used for sound classification (for example music versus voice) and for the discovery of sound events~\cite{lu_09}. The VibeFone application\cite{madan_06},
uses location, proximity, and tone of voice features to infer specific aspects of peoples' social lives. The mobile application has two special modes, the Jerk-o-Meter and the Wingman3G, in which VibeFone evaluates the subject's speech and provides feedback to subjects. Experiments have been conducted on several small scale data collections to measure and predict interest in conversation, and to measure attraction in a speed-dating scenario.

Other previous works in Reality Mining address a wide range of topics as follows. Wang et al.~\cite{wang_science09} model the mobility of mobile phone users to study the spreading patterns characterizing a mobile virus outbreak. They consider both location and proximity mobile phone data. They find that Bluetooth viruses spread slowly due to human mobility; however, viruses utilizing  multimedia messaging services could infect all users in hours. In~\cite{eagle2010network} individual calling diversity is used to explain the economic development of cities. Eagle et al.~\cite{eagle2010network} find that the diversity of individuals' relationships is strongly correlated with the economic development of communities. CenceMe~\cite{miluzzo_07cenceme} is a personal sensing system that enables activity sharing sensed automatically by mobile phones in a user's online social network. The sensed activities, referred to as ``sensing presence'', captures a users' status in terms of activities (e.g., sitting, walking), disposition (e.g., happy, sad), habits (e.g., at the gym, at work), and surroundings (e.g., noisy). These features can then be shared in popular social networking sites such as Facebook, Myspace, as well as instant messaging tools such as Skype and Pidgin. Wesolowski and Eagle~\cite{wesolowski_10} use mobile call logs collected over a one year period to better understand one of the largest slums, Kibera, located in Nairobi, Kenya. Additionally, individual calling diversity has been used to explain the economic development of cities in~\cite{eagle2010network}.

There is an increasing number of works on very large-scale data collections. The dataset used by Gonzalez et al.~\cite{gonzalez_08}, is from a phone operator, with the drawback of containing location information only when phone communication is present. They used mobile phone data to study the trajectories of human mobility patterns, and found that human trajectories show a high degree of temporal and spatial regularity, more specifically, that individual travel patterns can ``collapse" into a single spatial probability distribution showing that humans follow simple, reproducible patterns. The dataset contained 100 000 users over a period of six months. In~\cite{candia_08}, phone call data has been used to study the mean collective behavior of humans at large scale, focusing on the occurrence of anomalous events. The authors also investigate patterns of calling activity at the individual level and model the individual calling patterns (time between phone calls) as heavy tailed. The most recent work considering a very large scale mobility dataset obtained upon phone call initiation is by Phithakkitnukoon et al. \cite{phithakkitnukoon_12}, where they study the correlation between weather patterns and mobile phone usage.

Some state-of-the-art data collection campaigns include the Nokia-Idiap collection~\cite{kiukkonen_10}, which contains highly multimodal data on a large-scale of heterogeneous participants, consisting of family and friends, involving over 170 participants over a year of time. The data collected by Madan et al.~\cite{madan_11PC} occurs over a short duration with on the order of 70 participants, however, they target specific computational social science questions during the collection, which includes the collection of detailed questionnaires and surveys from the participants. The three main motivations are human political opinions, human obesity patterns, and human health including factors such as flu symptoms and depression.

\subsection{Exploiting Social Context}

One of the consequences of success of PUC is the introduction of computing applications which are based on social sciences. According to Tirri \cite{Tirri2010} pervasive communication technology together with sensor technologies is on its way to fundamentally change, beside other domains, social fabric of societies. PUC provides the infrastructure to sense the environment and equips the user to interact with it seamlessly \cite{Markarian2006}. To satisfy this vision, Weiser has already in 1999 recommended that pervasive computing solutions should also be unobtrusive and transparently integrated into social behaviour \cite{Weiser1999} \cite{Kumar2006}.

Sensing and interaction with the environment does not only involve infrastructure elements such as digital signs (electronic displays) \cite{Exeler2009}, interactive walls \cite{Ferscha2002},\cite{ferschaindividual}, \cite{Soro2008} and smart floors \cite{Rimminen2010}, \cite{Helal2005}, etc., but, to apply user-adaptive or context-aware behavior, also the users themselves \cite{Park2004}, \cite{Krejcar2010}, \cite{Baldauf2007}, \cite{Malatras2009}. Since in many cases a user (agent) is more than a digital device or entity, e.\,g. a human being, the inclusion of social behavior into pervasive applications is increasingly gaining importance. The collective paradigm, derived from pervasive computing, social media, social networking, social signal processing, etc., has recently become known as ``pervasive social computing'' \cite{Zhou2010}.

Recent developments within body worn sensors \cite{Haslgrubler2010}, \cite{Wirz2010}, \cite{Amft2006},\cite{farrahi2012socio} and ambient intelligence \cite{Treur2008}, \cite{McIlwraith2010} provide new possibilities to contribute to sensing of physical as well as cognitive/social attributes of human being, and moreover to integrate these into pervasive applications serving a smart environment \cite{Nakashima2010}. The wearable systems to sense the physical characteristics such as presence, location \cite{ferscha2009lifebelt}, locomotion (e.\,g. direction and speed of arm or leg motion) \cite{hoelzl2012locomotion} and body postures (e.\,g. sitting, standing and activity recognition) are already well developed by using accelerometers, gyroscopes, compasses and positioning/orientation sensors \cite{Wirz2010a,Wirz2009}, \cite{Lukowicz2010}, \cite{Kunze2010}. The new generation of wearable systems which could --for the first time-- measure the cognitive aspects (e.\,g. tension, happiness, excitement, etc.) \cite{Matthews2007} is gaining popularity. Examples of these sensors are EOG \cite{Bulling2009}, EEG \cite{Erdogmus2005}, and ECG sensors \cite{Riener2009:AutoUI:HeartOnTheRoad} as well as galvanic skin response (GSR) sensors or pupil diameter variation sensing \cite{Dufresne2010}. Similarly the development in ambient sensors to recognize the physical as well as cognitive aspects has progressed well beyond the video and audio streams analysis and has entered into implicit interaction paradigm \cite{Riener2008:ISVCS:DriverIdentSittingPostures}, \cite{Riener2011:IGIGlobal:ContinuousAuthentication}. General purpose sensing architectures have been developed, serving multi-purpose, multi-sensor, spontaneous and opportunistic sensing missions \cite{kurz2011opportunity}, \cite{kurz2011real}, \cite{kurz2012dynamic}, \cite{hoelzl2012goal} .

In many of these smart environments the ultimate beneficiary are humans incorporated as social individuals. A prerequisite for the successful application of personalized services (allowing contextualization on single person granularity) is efficient, safe, and unobtrusive user identification and profiling. According to Pour \cite[p.\,42]{Pour2008}, RFID can be good solution for pervasive identification respecting privacy/security of people, the usage of mobile phones could be, after unification of interfaces, another promising approach. Despite efficient available solutions, biometrics seems the inevitable part of the future of identification as it is unique, portable, and always with the user.

In addition to a profile which defines a user, the context also involves social relationships which can be woven deep down into a profile (e.\,g. family members, office colleagues, relationship status, friends, etc.) or can be formulated on the fly (e.\,g. passengers traveling in same train carriage, fans visiting a soccer game or drivers stucking in traffic jam). Overlying, the social relation of a person is the social behavior composed of individual preferences and the collective situation.

For example, if a user is getting out of a railway station in a hurry and pushing hard within a crowd, he may be getting late for work or there is an unusual situation e.g. a mass panic \cite{ferscha2010lifebelt}. A pervasive application designed to assist such a user should be aware of (i) user location, (ii) user state of arousal, (iii) user profile and (iv) environmental situation, to provide useful assistance. Knowing the location of user and the fact that he is excited, the context in which he is operating should be extracted either through usual user activity (his profile) or through situation sensing. The application should decide itself the reason for user arousal. Reaching to former reason has been a subject of context aware computing for couple of decades now. However, sensing and recognizing a crowd phenomenon is still a novice area of research due to complexity of involved social dimension. What complicate it further are human connections. For example the same user may act differently if he is accompanied with his child and perhaps he would require a different assistance from application.

Modeling a social system starts with modeling representative individual entities constituting such a system. These entities are heterogeneous with varying character and capabilities. In a social system we cannot model these entities at variable (using structural equations) or system (using differential equations) level. As an analytical method for social systems, the agent-based modeling is rapidly gaining popularity, due to its capability of directly representing individual entities and their interactions \cite{Gilbert2008}.

\subsection{Modeling Social Agents} \label{sec:social}

An agent based model provides appropriate agent level features that could define a social entity \cite{ferscha2011potential}. These features are: (i) Autonomy: ability to make its own decisions without a central controller, (ii) Social Ability: ability to interact with other agents, (iii) Reactivity: ability to react to a stimuli, and (iv) Proactivity: ability to pursue its goal on its own initiative. Each agent in the system may have its own version of implementing these features. Additionally, an agent based model allows multiple scales of social structures culminating naturally at a macro or societal level. None of other modeling approaches, for modeling a social system, comes as natural as agent based modeling approach. Formally, {\it agent-based modeling (ABM)} is a computational method that enables  to create, analyze, and experiment with models composed of agents that interact within an environment \cite{Gilbert2008}. Among the features which makes ABMs it an attractive choice for social modeling and simulation are:
\begin {itemize}
\item There can be one-to-one correspondence between real world actors and virtual agents which makes it easier and natural to design an agent and also to interpret the simulation results.
\item Possible heterogeneity in agents behavior advocate the usage of ABM in social systems.
\item It is possible to represent the space in which agents are acting directly into the ABM which makes modeling easier in an integrated environment.
\item Using ABM, agents can interact with each other at different granularities thus introducing the core social building blocks of communication and grouping etc.
\item ABM are able to implement learning / adaptation at local as well as global scale.
\item Many models implicitly assume that the individuals whom they model are rational. Herbert Simon \cite{Simon1955}, criticized this and suggested that people should be modeled as {\it boundedly rational}, i.e., as limited in their cognitive abilities and thus in the degree to which they are able to optimize their utility \cite{Kahneman2003}. ABM makes it easy to create boundedly rational agents. {\it In fact, the challenge is usually not to limit the rationality of agents but to extend their intelligence to the point where they could make decisions of the same sophistication as is commonplace among people \cite{Gilbert2008}}.
\end{itemize}

Among the major difficulties of large scale ABM simulations, on the other hand, are: (i) agents’ heterogeneity, (ii) overlapping granularity of interacting entities and (iii) complex space models. In the scientific computing / computational science community a generic term for second aspect is multiscale modeling which can be narrowed down as social organization in systems addressing social phenomena.

\noindent \textbf{Agents' Heterogeneity} In a large scale agent based system, the agents are typically heterogeneous in nature. There is a variety of behavior for each agent in chemical, biological, economic or engineering systems. However a social system addressing the cognitive aspects of participating entities (human beings) is much more complex. The following aspects highlight the challenges involved:
\begin {itemize}
\item {\bf Individualism:} Each individual agent can be as varied as physically and behaviorally different as the humans are.
\item {\bf Decision making:} The process of social decision making is not a simple one. It may involve unlimited options to explore. Even a single decision may involve complex formulations. It is not practical to formulate a complete rule set before hand. The decision making rules evolve all the time as an agent learns from previous decisions and decisions taken by others. Most decisions are not independent and depends on parameters from influencing entities.
\item {\bf Behavioral Adaptivity:} Behavioral adaptivity can be distinguished from behavioral learning as it targets the change in behavior due to dynamics of the other entities in interaction whereas learning describes the cause-affect relation which changes the rule-based with experience.
\end {itemize}

\noindent \textbf{MultiScale Modeling} Interaction modality (as described at agent level) is necessarily at a single scale. Multiscale modeling is the study of systems which operate at a multiple resolutions. One of the strengths of multiscale modeling is its ability to provide and link a system's functionality at different length vs. time scales \cite {Karl2010}. An example of such a system is discussed in \cite {Karl2010} in the domain of chemistry. The different levels of simulations (calculations) discussed are quantum mechanics calculations (within an atom), atomistic simulation (within atoms) and coarse-grained simulations (e.g. within molecules). As highlighted in the article, a major challenge for all multiscale simulations is to transfer the knowledge gained from one resolution to another.  Multscale resolution when applied in natural systems poses a major challenge due to emergence of macroscopic behavior (which is usually a subject of interest) due to microscopic interaction. The challenge arises from explicitly modeling the microscopic behavior which can vary for each of the interacting agent (as discussed in above sub-section) making simulation computationally infeasible to do sufficiently long simulations where macroscopic behavior emerges \cite{Crommelin2010}.

\noindent \textbf{Complexity of Space and Agents Clustering} The concept of space has two meanings; ontological and physical. In first, we take space as a room, or a specific corridor connecting two rooms. The conceptual definition of space in this way guarantees a more convenient behavioral analysis granularity focusing on a conceptual basis of analysis rather than unnecessary physical (coordinates etc.) details of a space. In second, we describe space in physical domain which is necessary for agent's behavioral implementation at agent level. Many observational evidence can only relate behavior of an individual on ontological level. For example an interviewee can only relate his experiences to a contextual space, or the observations reported by an ambient device may report a flow of people through an exit. This makes it difficult to extract the physical space from contextual space. Additionally a true representation of space in modeling also derives its complexity from complexity of the environment itself. Most real environments cannot easily be represented in a true-to-scale and representative digital clone.

Considering that an agent's specification includes the spatial aspects of the environment, the decision making of an agent in a social system is absolutely dependent on group size. Within a cluster it is not necessary that all the agents would / should be communicating with each other as agent's individualism describes the desire or capabilities to communicate / {\it interact}. Based on the influences from the group and agent's own expectations and experiences, an agent can perform an {\it action} or {\it adapt} its behavior. This simple specification of a social agent can intuitively be derived from ABM specifications discussed and can be used to model any social phenomena.

\subsection {Modelling Trust}

While exploring a common social attribute, Sissela Bok in her seminal work: \emph{Lying: Moral Choice in Private and Public Life} \cite{Bok1999}, signifies the importance of closely related issue of {\bf trust} in societies as \emph{... trust is a social good to be protected just as much as the air we breathe or the water we drink.} In Golembiewski and McConkie view, \emph{Perhaps there is no single variable which so throughly influences interpersonal and group behaviour as does trust ...} \cite{Golembiewski1988}, cemented further by Luhmann as \emph{Trust ... is a basic fact of human life} \cite{Luhmann1979}. Nevertheless, instead of indulging into the unlimited stretch of social, psychological, and even biological aspects of trust and related issues, we have focused on formalism of trust and related issues of cooperation and collaboration in agent-based systems; a technology designed and suited for mimicking a social interaction based system like societies of humans and devices. However, the essence of factors influencing the interaction and autonomous / distributed decision making in agent-based systems still inherits its foundation from understanding and hence borrowing the concepts from social, psychological and biological means wherever it applies.

In a socio-technical system, the basic purpose is to assist the user in decision-making. Since the introduction of the concept of pervasive computing, the social aspects of the society and its integration into decision making process has become absolutely necessary, but at the same time a challenging task. Whats makes it even more challenging is the variation in interaction modularities; for example, human-to-human, human-to-device, device-to-human and device-to-device interaction. Even if we ignore the almost impossible task of "precisely" quantifying a social aspect (for the purpose of using it it the system) for the time being, the mere challenge of dealing with quantifying a social aspect for these modularities is sufficiently tough. Nevertheless, a clear progress is underway primarily due to the development of appropriate modeling techniques that not only support the interactions between devices, humans and any of its combination, but also provide mechanisms similar to the decision-making mechanisms in human societies. One of the most promising technology in this regard is agent technology or an Agent-Based System (ABS).

What makes an ABS unique w.r.t a socio-technical system is its support for "human-like" behaviour and its capability of decision-making in a cooperative fashion. An agent's capabilities of {\it autonomy} and {\it intelligence} helps reaching to a rational decision whereas capabilities of {\it cooperation} and {\it distributed} interaction helps achieving society-level goals, where a society of agents comprises of a meaningful group of agents in a population, somehow related to each other. These capabilities help agents perform localized decision making without the knowledge of global conditions (mostly analogous to the human-decision making process). The social aspects affecting the individual attributes of an agent and hence its decision making capabilities are application specific. However, whatever the application is, the very fact that the agents are working together means that {\bf trust} is a relevant issue. With this report, we endeavor to understand the affects of involving trust as a social aspect affecting the decision-making process of an agent-based socio-technical system. Towards this effort, using agent based technology, a formalism of trust is also presented and used in a specific scenario.

The main work related with trust comes from the fields of sociology, (social) psychology, economics, biology and philosophy.
\\

\noindent \textbf{(Social) Psychology - Morton Deutsch}
Perhaps the most accepted definition of trust comes from Deutsch work from 1962 \cite{Deutsch1962}, which states that:
\emph {
\begin{itemize}
\item (a) the individual is confronted with an ambiguous path, a path that can lead to an event perceived to be beneficial ($ \ Va^+ \ $) or to an event perceived to be harmful ($ \ Va^- \ $);
\item (b) he perceives that the occurrence of $ \ Va^+ \ $ or $ \ Va^- \ $ is contingent on the behaviour of another person; and
\item (c) he perceives the strength of $ \ Va^- \ $ to be greater than the strength of $ \ Va^+ \ $.
\end{itemize}
If he chooses to take an ambiguous path with such properties, I shall say he makes a trusting choice; if he chooses not to take the path, he makes a distrustful choice.
}

The usage of perception suggest that trust is {\it subjective}. It means that it can vary from one individual to the other, even if observable social aspect is the same. Implicitly, within the definition, there is an "cost-benefit" analysis. In many theories related to human decision making, there is a tendency of theologists to explain the behaviour using cost-benefit (or utility) analysis, as done by Deutsch. However, in practical terms, human as well as a computing system cannot spend an unlimited time on this analysis \cite{Good2000}. The "intelligent" guessing is always adopted by humans due to limited time in hand or sometimes laziness. This should also be the case with ABS, as computing resources would never be enough to exhaust all possibilities. The cost-benefit analysis, whether exhaustive or guessed, is always performed in certain circumstances which determines the type of trust a person is willing or forced to adopt.


Different types of trust are possible primarily based on the circumstances and individual personality (Circumstantial Trust). Deutsch explains these types with the help of a story: \emph{The Lady or the Lion}. We have extended the consequences drawn from the story to explain the types of trust wherever it is necessary. The story goes on like this: {\it There was a princess who has a suitor. The King knew about this relation and was furious. He ordered suitor into a pit which has two exits. Behind one exit there was a hungry lion waiting for the prey. Behind the other exit, there was a beautiful lady, presented as a replacement of the princess. The suitor had to make a decision knowing the two options but not exactly; behind which exit, which of his possible fate was waiting. He was also instructed to choose an exit, otherwise he would be executed. Before he made a choice, he saw princess pointing towards one of the exits.} The rest is left for the imagination of the reader.
\begin{itemize}
\item {\bf Trust as despair:} The very fact that the suitor would be executed if he does not proceed with an exit, explains trusting a choice (making a decision) in despair, without considering the fact whether suitor follows the instruction from princess or not.
\item {\bf Trust as social conformity:} The decision of suitor to follow the instruction from princess or not, presumably in addition to other factors, depends on social norm set by the society. In this case, a social norm of 'woman's gentleness' may be prevalent, which may suggest to the suitor that a woman would always save life of her lover, even if he falls into the arms of another woman.
\item {\bf Trust as virtue:} The above argument is also true in this case. Additionally, presuming that the princess is following the social norm or virtue pointing him towards other woman, he may follow the other exit just to save his virtue.
\item {\bf Trust as masochism:} The difference between choice made based on virtue and masochism is the state of disappointment which forces the suitor to face the lion.
\item {\bf Trust as faith:} Even knowing that the princess would be pointing towards the exit having life, the suitor may follow the other exit having a faith that he would kill the lion, and would never betray princess.
\item {\bf Trust as innocence:} Imagine that the suitor never knew what lies behind the exits. He would make a choice; either a wild guess (in the absence of princess instruction), or a mindless following (in the presence of princess instruction).
\item {\bf Trust as Risk taking / gambling:} Opposed to the name given to this category, this category represents a rational behaviour given that the suitor is not influenced by emotional and normative attributes. In this case, the suitor would weigh the potential gains of wining the lady against the potential losses of losing to lion. If he thinks that there is no way he could beat the lion, he may follow the exit leading to the lady, even if he did not want to betray the princess. And there are many more combinations which could be imagined. Another aspect of gambling can be that the suitor knows that the princess would let him die rather then letting him live with another woman; in which case he may gamble against the princess's instruction.
\item {\bf Trust as impulsiveness:} This is similar to the Prisoner's dilemma in which a retreat in current move may guarantee a gain in future. Even knowing that there is more risk in facing the lion, the suitor may proceed towards that exit knowing that if he beats the lion, the princess would be with him. And also knowing that if he chooses the other option, he would be killed anyway; this time by princess.
\item {\bf Trust as confidence:} This is a more deterministic variation of gambling. In fact, the confidence about one of the choices is determined by many factors (probably all above factors). A person chooses an option in which he is most confident of utilizing all possible cost-benefit measures. That is the reason, Deutsch chooses 'confidence' as a type of trust he would be focusing on.  Considering the many dimensions a confidence level is based on, Deutsch coins the term {\it Utility} to encompass everything into a function. The more the utility of a choice is (a rational choice, not necessarily unemotional), the more confident the user is about it, thus trusting it more.
\end{itemize}


Deutsch's work is guided by the following hypotheses:

\begin{itemize}
\item {\bf Hypothesis 1:} \emph{Given that $ \ Va^- \ $ is stronger than $ \ Va^+ \ $, a trusting choice will occur if: $ \ Va^+ *  S.P.^+ > Va^- *  S.P.^-  + K \ $}. Where $ \ S.P.^+ \ $ is the subjective probability of attaining $ \ Va^+ \ $, and $ \ S.P.^- \ $ is the subjective probability of attaining $ \ Va^- \ $, and K is the constant called "security level". A cost-benefit analysis.
\item {\bf Hypothesis 2:} \emph{The more remote in time the possible occurrence of $ \ Va^- \ $ as compared with that of $ \ Va^+ \ $, the more likely is that a trusting choice will be made.} The concept of memory.
\end{itemize}

\noindent \textbf{Sociology - Niklas Luhmann} Luhmann's main thesis is that trust is a means for reducing the complexity of the society \cite{Luhmann1979}. Instead of utility analysis, he relates trust with risk analysis, and states: \emph{It becomes ever more typical and understandable that decisions cannot avoid risk.} Indeed, \emph{Trust ... presupposes a situation of risk.} \cite{Luhmann.2000} Realization and assessment of risk does not make life any harder. In fact, without realizing and dealing with risk, it would be impossible to live a competitive and useful life due to the fact that life is too complex and short for rational manipulations.
\\

\noindent \textbf{Sociology - Bernhard Barber} In his 1983 book, \emph{Logic and Limits of Trust} \cite{Barber1983}, Bernhard Barber attempts to clarify the concept of trust. He relates trust with the social relations and equate it with expectations drawn from experiences within social relationship.
\\

\noindent \textbf{Can We Trust Trust? - Diego Gambetta} With the deliberations from Diego Gambetta \cite{Gambetta2000}, we will try to conclude the discussion about definition of "trust". In his article "Can We Trust Trust?", Gambetta states that \emph{Trust is a peculiar belief predicated not on evidence but on the lack of contrary evidence - a feature that ... makes it vulnerable to deliberate destruction.} Ignoring the second half of the sentence, he extends the concept of Luhmann, trying to reduce the complexity involved in cost-benefit analysis or tracking the social relationship history. He simplifies the trust relation as being trustworthy unless we are actually confronted with evidence which goes against it. In the following, we provide a review of Gambetta thesis.

\emph{... two general reasons why - even in the absence of 'thick' trust - it may be rational to trust trust and distrust distrust ... The first is that if we do not, we shall never find out: trust begins with keeping oneself open to evidence, acting as if one trusted, at least until more stable beliefs can be established on the basis of further information. The second is that trust is not a resource that is depleted through use; on the contrary, the more there
is the more there is likely to be ... }

\emph {... if behaviour spreads through learning and imitation, then sustained distrust can only lead to further distrust. Trust, even if always misplaced, can never do worse than that, and the expectation that it might do at least marginally better is therefore plausible. However, while the
previous reasons can motivate rational individuals to trust - at least to trust trust - this reason alone cannot, for though everyone may concede it, if the risk of misplacing trust is reputed to be high, no one wants to be the first to take it. It is enough, however, to motivate the
search for social arrangements that may provide incentives for people to take risks.}

This explains why people tend to trust rather than distrust as a default behaviour.

\section{Laying Ground for  Socio-Inspired ICT}
\label{intro}

Advances and the confluence of contributions in core areas of computer science - like artificial intelligence, natural language processing, computer vision, robotics, machine learning, and cognitive science - have greatly contributed to what we observe to be an ICT exhibing characteristic principles of awareness. The development of more human-like computing capabilities has been the focus of several research groups in academia. However, one distinguishing feature has not been considered in much detail up until now. Computers completely lack human social capabilities. Our human social systems play critical roles in individual cognition, intelligent, and progress. In order for computers to better understand our world, particularly in a more human-like manner, one potential mostly unexplored direction is with respect to human social capabilities. Social capacity is one additional dimension of intelligence and would lead to even more intelligent computing, better usability of existing technologies, and new applications and technologies. To quote Pentland, ``ultimately [today's communication technologies] fail because they ignore the core problem: Computers are socially ignorant``~\cite{pentland_05socially}.

Traditionally, human social behavior has not been fully understood due to two critical factors. Firstly, humans both as individuals and groups are complex in nature. The complexity stems from differences between individuals, as well as the multiple factors which may be relevent, such as context, environment, and even the interacting individual, group size, and or seasonal effects. Secondly, large-scale social behavior-specific data for scientific analysis is difficult to obtain. With the emergence of new forms of large-scale human behavioral sensing techniques, such as reality mining~\cite{realitymining_website,eagle_09} and computational social science (CSS)~\cite{lazer_09}, a better understanding of human social behavior at larger scales is becoming a possibility. We are at the point where ``big data'' can help research efforts to address previously unexplored directions in computer science, particularly in an interdisciplinary manner with the social sciences.

There is an increasingly large body of research in computer science relating to human social behavior sensing, social signal processing, social awareness, social intelligence, and socio-inspired ICT. These research topics, some of which have gained more momentum than others, are all centered around human social behavior and its detection, understanding, and simulation. A very broad range of research in psychology, sociology, economics, and physics, in addition to computer science, present research on topics which are relevent. Next we define socio-inspired ICT as the transfer or automation of human social behavior understanding to ICT systems and identify several research directions from which socio-inspired ICT could greatly benefit and build upon.

\subsection{Social Signals and Social Architectures}

Social signals have recently been defined as a communicative or informative signal that, either directly or indirectly conveys information about social actions, social interactions, social emotions, social attitudes, and social relationships~\cite{poggi_10}. Informative signals are distinguished from communicative signals based on the intention of the signal producer. If there was no direct intention of conveying a particular social information to the receiver, the signal is considered to be informative and in contrast, the signal is communicative if the emitter produces a signal with the goal of having the receiver obtain some belief~\cite{poggi_10}. Further, what characterizes ``social signals'' as social is not that they convey information from one entity to another, but that their ``object'', the type of information they convey, is social~\cite{poggi_10}.

While we agree with the definition of Poggi and D'Errico~\cite{poggi_10}, we believe this definition can be stated in such a way that social actions, social interactions, social emotions, social attitudes, and social relationships need not be differentiated from actions, interactions, emotions, attitudes, and relationships. To get a better insight into what social signals are and are not, we should additionally consider the term social behavior from the perspective of the definition in physiology and sociology. {\bf Social behavior} is behavior directed towards society, or taking place between, members of the same species~\cite{wikipedia_socialbehavior}. Given the previous definition of social signals~\cite{poggi_10}, this would imply a person is displaying social signals with their pet, though this would not traditionally be accepted as social behavior as it is not considered true that people socialize with their dog, for example, even if they send them many emotional displays of affection. Additionally, it
is not clear how anti-social signals could be differentiated from social signals. We believe the definition of social signals should additionally consider the  purpose of the exchange, which is taking the interests, intentions, or needs of others into account.

We build on the previous definition as follows. {\bf Social signals} are a communicative or informative signal that either directly or indirectly conveys information about the actions, interactions, emotions, attitudes, and relationships, and necessarily takes into account the interests, intentions, or needs of others in society, or between members of the same species. This would be in contrast to anti-social signals, which would not take the interests, intentions, or needs of others into account. By taking the intention of the signal into account we remove the need to differentiate between social interactions and interactions, and so on.
{\bf Socio-inspired ICT} now is the recognition or automation of aspects relating to human social behavior to ICT systems.  We identify two types of \textbf{Social Architectures} under which socio-inspired ICT research fits, the differentiation stems from the recognition versus automation goals of the ICT system.
 {\bf Social Artifical Intelligence} The first type seeks to 'encode' human social behavior into computing. Artificial intelligence (AI) is the intelligence of machines, and the branch of computer science that aims to create it. The methodology behind AI is the understanding of human intelligence, how it works, and transferring that knowledge to computing. Similarly, the first architecure, which we call social artificial intelligence, transfers the understanding of human social behavior to computers for varying purposes, resulting in social AI. Social AI architectures can be used for improved human-computer interaction. For example, robots with human social capabilities for improved learning and development functionality, or new advancements in gaming technologies. Imagine a network of agents which human social behavior. Massive socio-inspired agent-based models would fall into this category. Replacing agents by computers, or even cars to have a network of social computers or social cars; these social
AI networks could lead to many future research directions.
{\bf Social Signal Recognition} The second type of architecture relating to socio-inspired ICT, under which most current research can be categorized, is that where human social signals are sensed, recognized, interpreted, and processed by computers, though they do not try to simulate this behavior. Research in this category includes computing equipped with varying forms of sensors which can automatically and subjectively detect, collect, process, and correctly infer social signals.

\subsection{Cognition and Social Behavior:  Attention and Perception}

Cognitive sciences typically de-emphasize and often exclude social and emotional factors from their studies, and have been criticized for neglecting the roles of emotions and humans as inherently social beings~\cite{stanford_encyclphilos}. The affective sciences consider human emotions in full detail~\cite{picard_97}. A complete understanding of human cognition is not necessary in order to develop socio-inspired ICT, though there are several previous works in the cognitive sciences which are very relevent and the development of more advanced and complex socio-inspired ICT will likely benefit greatly from advancements in cognitive science. As aspects of critical importance in understanding social behaviour, and with that the principles of social interaction we highlight human attention and perception. Human attention is the first source of perception, consequently als awareness towards information and other individuals. Over the last decades, attention research has succeeded in identifying several attention types as well as physiological mechanisms and neural processes and revealing its relation to memory, learning, decision making and perception. In the history of attentional research many different attentional mechanisms have been discovered and according descriptive models have been developed. Usually, these models adequately describe single or several aspects of attention, e.g. the ambiguity of single- and multi-tasking capability, whereas a general, overall theory of attention is still missing.
Attention research has a tradition going back to James \cite{ebalp:James1880} in the late 19th century. In the first half of the 20th century, attention has been disregarded as being 'not measurable' and it needed the First Cognitive Revolution in the middle of the 20th century for attentional research to be presentable again.

\textbf{Single Channel Theory (SCT)}
The fundamental issue of attention research at its early stages was the parallelism of attentional processing. This comes down to the question whether human brain is capable of actually processing different tasks at the same time, effectively carrying out multi-tasking or if single stimuli are selected for orienting of attentional and these tasks then are processed serially. To investigate competitive selection processes, Craik \cite{CRAIK.1947} and especially Broadbent \cite{Broadbent.1969} conducted dichotic listening experiments during which they overloaded subjects with information from different sources at the same time. As a result of his experiments, Broadbent interpreted the attentional process an information processing system, which is equipped with filter mechanisms that separate relevant from irrelevant information. He concluded that humans are only capable of orienting their attention towards a single channel at a time, hence, tasks are always processed serially. In technical terms, this approach resembles a single core processor, performing scheduling processes to ensure real-time capability.

\textbf{Early vs. Late Selection}
In the ongoing, several alternative versions of the SCT evolved usually differing in less restrictive modifications of the filter mechanisms and moment of selection. Treisman \cite{Treisman.1980} proposed a more flexible filter mechanism in her \textit{Feature Integration Theory}, in which disregarded stimuli are merely attenuated and not completely blocked as in Broadbent's model. Deutsch \& Deutsch \cite{DEUTSCH} stated that the actual filtering happens at a late processing stage and all input stimuli are processed equivalently until entering short-term memory stage. Lavie \cite{Lavie.2004} combined both approaches by connecting the moment of selection to mental workload in her \textit{Perceptual Load Theory}. According to her findings, early selection is carried out in case of high workload whereas in case of low workload, selection happens at a late stage.

\textbf{Capacity Theory}
In contrast to SCT, the \textit{Capacity Theory} (CT) assumes that human attention is only limited by an overall capacity of attentional resources and which can be distributed among different tasks. Following this principle, individuals are only capable of handling a certain workload, independent of dealing with single or multiple channels. CT is based on the observation that tasks can be carried out simultaneously as long as they are sufficiently automated and do not require high mental effort. The assumption that attention is depending on the overall workload is supported by experiments from Bahrick and Shelley \cite{BAHRICK.1958}, Schneider and Shiffrin \cite{Schneider.1977} and Damos \cite{Damos.1978}, in which training enhanced the performance of subjects in multi-tasking applications by lowering the demand for resources with growing automation. This is supported by findings from Seligmann et al. \cite{Seligmann} and Woollacott and Shumway-Cook \cite{Woollacott.2002} who found that usually highly automated tasks like human gait demand significant resources in case of gait disturbances. Hence, attention resources are regarded as constant and limited but freely distributable. In contrast, Kahneman \cite{Kahneman.1973} proposed a dynamic amount of capacity, depending on the amount of current arousal. He identifies \textit{Mental Effort} as the major control component of resource allocation, being directly proportional to arousal and hence for mental resource management. This implies, that for complicated and important tasks more resources are allocated automatically, but the voluntary control over arousal and effort is severely limited. Sarter \cite{Sarter.2006} describes the link between effort and arousal as the 'motivated activation of attentional systems' and developed a neuronal model which successfully connects  motivation processes to neuronal activations. Still, attentional effort remains a rather neglected aspect of attention research.

\textbf{Multiple Resource Theory}
SCT and CT cover different aspects of competitive selection depending on the task and general situation of the subject. A more general attention model has evolved in the last decades which succeeds in integrating these different attentional aspects into a single model. Wickens 
\cite{Wickens.2008b} elaborated the approach of \textit{Automaticity} to a more general model of \textit{Multiple Resource Theory} in which tasks can be carried out simultaneously as long as they differ in their type of resource demand. He proposes a four dimensional resource model  in which he distinguishes between perceptual modalities, processing stages, visual channels and processing codes. This implies, that simultaneous performance of a visual and an auditory task causes less interference of allocated resources than performing two tasks involving related missions e.g. visual search. This model includes findings from the CT and also implies Single-Channel phenomena for tasks with a similar resource allocation. The concept of Effort is considered as a factor on filtering and selection which will be addressed in the ongoing.
The history of attention research has brought forth diverse approaches and models. Whereas early theories are restricted to the description of isolated attentional mechanisms, current approaches based on and refining the \textit{Multiple Resource Theory} show better overall applicability. This theory best describes the complexity of the human attentional system and is furthermore suitable for application in general and technical use-cases.

\textbf{Attention and Awareness}
Awareness of activities is generally defined as processes which can be reported either verbally or via actions \cite{Posner.1980b}, \cite{mgarces:dijksterhuis_goals_2010}. The question of human awareness is closely related to the confrontations between overt and covert attention and attention control via top-down and bottom-up processes. To some degree, it investigates whether overt and covert attention are oriented at the same destination. Moreover, it covers the question whether attention processes which have been triggered unconsciously and involuntarily by external stimuli (bottom-up processes) reach the actual awareness level or stay below the consciousness threshold. Attention and Awareness are often used interchangeably, or, stated to be linked to an inseparable degree \cite{WCS:WCS27}. Others soften the constrictions to the assumption that there is no attention without awareness \cite{Taylor.2003}. This assumption has been disproved in recent research by Lamme  \cite{Lamme.2004} and Koch \cite{Koch.2012} who state that the two phenomena are closely related but not necessarily connected. Already in  1973, Kahneman \cite{Kahneman.1973} integrated a bypass option for the stage of conscious perception. The automatic capturing of attention by exogenous cues without necessarily triggering awareness is investigated by McCormick \cite{McCormick_1997}.  Duque et al. \cite{Fernandez-Duque:2003:RCS:1162383.1162384} worked on computing neuronal measurements to differ awareness and attention by electro-physiological signals. Brown and Ryan \cite{Brown_Ryan_2003} created the Mindful Attention Awareness Scale (MAAS), a psychological awareness scale to model impact of mindfulness on well-being. MacKillop and Anderson \cite{MacKillop_Anderson_2007} and Dam et al. \cite{VanDam_Earleywine_Borders_2010} successfully validated this MAAS scale using different methodologies.

\textbf{Intentions, Goals and Plans}
Human behavior is motivated by goals and plans, with plans referring to conscious intentions while goals can exist at both levels of consciousness. These goals and plans are equipped with a priority attribute, which indicates how targeted the fulfillment of this goal will be pursued and how easily people can be distracted. Bisley and Goldberg \cite{bisley} explored the aspect of intention and its physiological representation in the parietal lobe to infer on priority creation in attention and selection. They succeeded in identifying the lateral intraparietal area working as a priority map and being controlled by top-down and bottom-up processes. Dijksterhuis et al. \cite{mgarces:dijksterhuis_goals_2010} use the term 'goals' instead of 'intention' and assume that these are major top-down components that drive attention. Both of them are not necessarily connected to awareness. They state that, `goals guide behavior through attention, and this guidance can occur outside of a person's awareness'. The fact that goals can be imposed unconsciously is supported by experiments of Bongers et al. \cite{Bongers2009468}, additionally indicating a negative influence on perceived self-esteem in case of failure, even on unconsciously activated goals. Okuda et al. \cite{Okuda.2011} discovered an automatic regulation of attention between task performance and future goals without any intervention of conscious control systems, indicating how even task management can be carried out automatically at an unconscious level.

\textbf{Emotions and Instincts}
Emotions and instincts represent abstract forms of build-in goals and plans which are only directed at unconscious and automated processing. Bradley \cite{PSYP:PSYP702} concentrated on the most fundamental motivational system which is survival instinct, and identified mechanisms that affect orientation of attention and behavior. Similar to Bradley, Ohman \cite{Ohman.2001} investigated how strong emotions influence automatic attention capture, and found that evolutionary relevant threatening stimuli automatically triggered fear, which supported higher arousal states and higher priorities in the selection process. In her extensive review about the effects of emotion on attention, Yiend \cite{Yiend.2010} states, that there is no general \textit{pop-out effect} of negative information, but the visual search for negative or threatening information runs much faster. This supports assumptions from capacity-based attention models in which additional resources can be allocated in case of states of high arousal. Phelps et al. \cite{Phelps.2006} found evidence that emotion facilitates early vision, and vision is improved in the presence of emotional stimuli. Dolan \cite{Dolan.2002} describes how emotion influences decision-making processes by relating emotions from past decisions to future determinations, thus also connecting learned experiences to future decision making and behavior.

\textbf{Perception}
In order to receive information from the environment the sensory organs such as the eyes, ears, and nose, receive inputs and transmits information to the brain as part of a sensory system~\cite{mcleod_07}. A theoretical issue on which psychologists are divided is the extent to which perception relies directly on the information present in the stimulus. Some argue that perceptual processes are not direct, but depend on the perceiver's expectations and previous knowledge as well as the information available in the stimulus itself~\cite{mcleod_07}. Two proposed processes in perception have theorized about this issue, the first is a direct theory of perception which is a 'bottom-up' theory ~\cite{Gibson_66} and the second is a constructivist (indirect) theory of perception and is a 'top-down' theory~\cite{gregory_70}. Bottom-up processing~\cite{Gibson_66} is also known as data-driven processing, because perception begins with the stimulus itself. Processing is carried out in one direction from the retina to the
visual cortex, with each successive stage in the visual pathway carrying out ever more complex analysis of the input~\cite{mcleod_07}. Top-down processing~\cite{gregory_70} refers to the use of contextual information in pattern recognition. For example, understanding difficult handwriting is easier when reading complete sentences than when reading single and isolated words. This is because the meaning of the surrounding words provide a context to aid understanding~\cite{mcleod_07}.

From an ICT perspective, machine perception had previously been proposed as a promising future technology from a human-computer interaction perspective~\cite{crowley_00,turk_00}. Along the same lines, perceptual user interfaces (PUIs) are designed to make human-computer interaction more like how people interact with each other in the real world~\cite{turk_00}.
The chameleon effect is the nonconscious mimicry of the postures, mannerisms, facial expressions, and other behaviors of one's interaction partners, such that one's behavior passively or unintentionally changes to match that of others in one's current social environment~\cite{chartrand_99}. A perception-behavior link has been suggested as the mechanism involved, whereby the mere perception of another's behavior automatically increases the likelihood of engaging in that behavior oneself~\cite{chartrand_99}.

\section{Concluding Remarks}

An analsis of the evolution of ``aware ICT'' over the past two decades,
together with a systematic consultation of the scientific community
and a review of the literature revealed needs for, features of and even
first concepts towards what we call socio-inspired ICT in this position
paper. Current literature addresses research issues ranging from
situation, to activity and even cognitive state (e.g. emotion), identifying
needs to also address social capacities and bahaviours of individuals
when designing, developing and deploying novel ICT.

\begin{figure}[h]
  	\centering
	\includegraphics[width=0.9\textwidth]{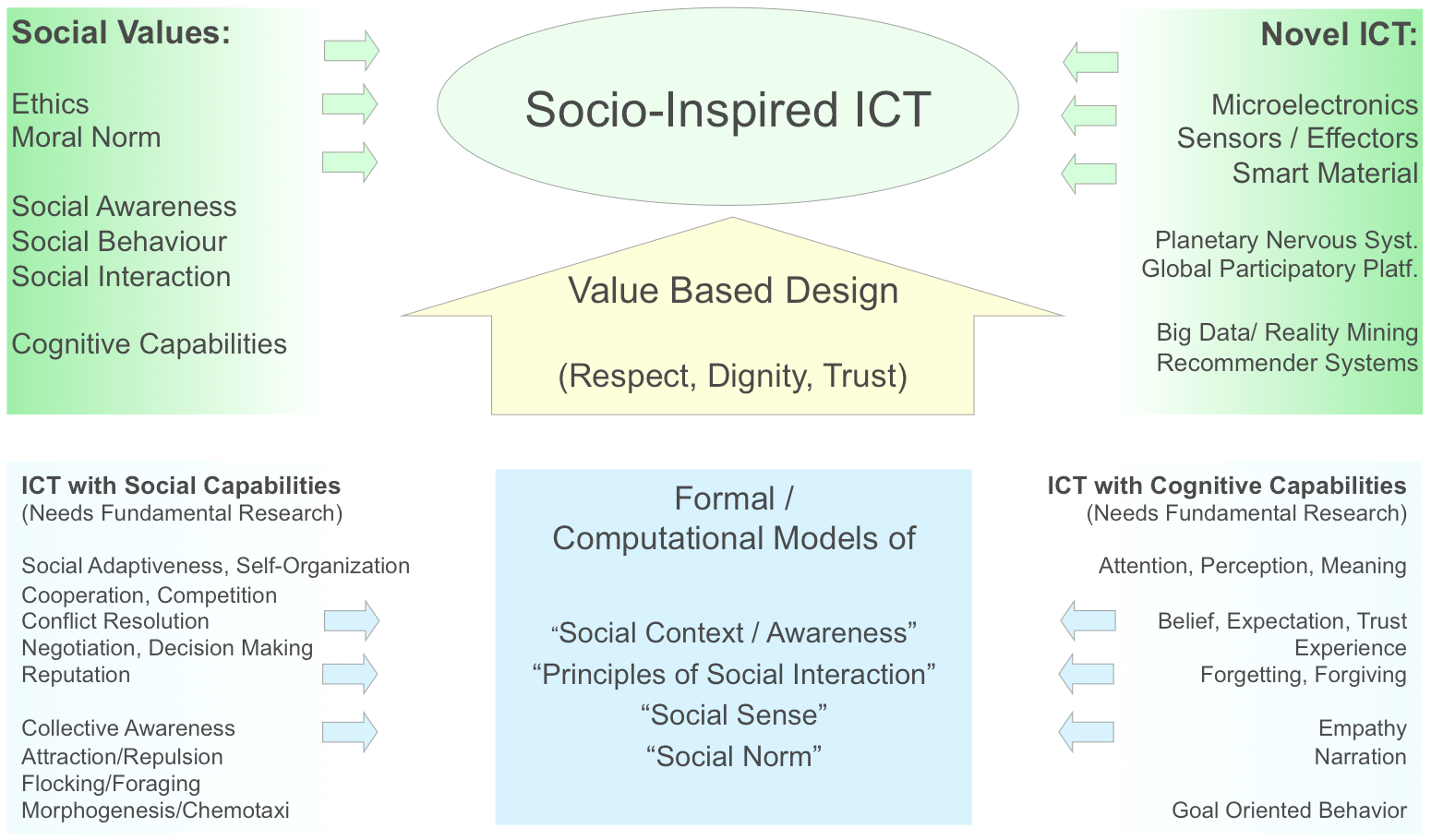}
	\caption{Research Questions Situated around Socio-InspiredICD}
   \label{fig:socioinspiredICT}
\end{figure}

In order to lay ground for the evolution future ICT systems, the understanding
of principles of social engagement, experiencing and behavioral consequences,
as well as the respect for individual and social values (``value based design'', i.e.
ICT designs respecting values like trust, souveregnity, respect, dignitiy, privacy, etc.,
see Fig. \ref{fig:socioinspiredICT}) reveal as an apparent, prevalent claim throughout
many disciplines potentially fertilizing and steering this evolution  (Computer Science,
Social and Cognitive Sciences, Complexity Science). A  "framework of principles" for
socially adaptive ICTs is in demand, formalizing the process on how individuals engage
in social activities and experience social relations, and make it the design-,
implementation-, and operational principles of forthcoming large-scale ICT systems.
Towards formal and computational models underpinning socio-inspired ICT,
multi-scale (in time and space) sensing system able to capture, analyse and store social
engagements, impressions, activities and behaviors are needed. Concepts like
``social context'', ``social awareness'' and ``social sense'' need to be studied and
formalized both at the level od individual social beings, as well as at the level of
groups, collectives, and even whole societies.  Towards this, the potentials of sensor systems
needed and available to capture impressions of social interactions have to be investigated,
structuring sensor technology with respect to physical constraints (size, weight, etc.),
mobility, energy consumptions, availability and fault tolerance, quality of service,
accessibility and seamless integration. The methodological and algorithmic integration
of the foundational principles of human social capabilities as well as human cognitive
capabilities will ultimately allow for design-, development- and operational principles
of novel ICT with social capabilities as well as cognitive capabilities. This ``alignment'' of
future generation ICTs with the capacities and capabilities of individuals and societies
gives rise for a flourishing symbiosis among society and ICT at the confluence of social
values and technological progress.

The provision of  a formal, computational framework
of consolidated principles and models of $(i)$ individual socio-cognitive capacities
and $(ii)$ collective social capacities appears to be the critcal precondition for
a successful development of a reference architectures of globe spanning, participative,
trustworthy, engaging, socially adaptive ICTs that are $(i)$ value sensitive by design,
that $(ii)$ adapt to and co-evolve with the dynamics of the norms and values of societies,
and $(iii)$ center on a long-term stability of humankind.

\nocite{dash_03,rahman_11,cohen_03,economist_jun132011,ostrom_90,kesting_08,kesting_10,lammer_08,helbing_12,carvalho_12}

\bibliographystyle{spmpsci}
\bibliography{references,references2011,ssp_refs,ubicomp,attention}
\end{document}